\documentclass[floatfix,twocolumn,prb]{revtex4}
\usepackage{subfigure}
\usepackage{color}
\usepackage{multirow}
\usepackage{epsfig}
\usepackage{graphicx}
\usepackage{amsmath}
\usepackage{amssymb}
\usepackage{bm}
\usepackage{dcolumn}
\usepackage{braket}

\begin{document}

\title{Anisotropic intrinsic anomalous Hall effect in ordered $3d$Pt alloys}

\author{Hongbin~Zhang}
\email[corresp.\ author: ]{h.zhang@fz-juelich.de}
\author{Stefan Bl\"ugel}
\author{Yuriy~Mokrousov}
\affiliation{Peter Gr\"unberg Institut and Institute for Advanced Simulation, 
Forschungszentrum J\"ulich and JARA, D-52425 J\"ulich, Germany}

\date{\today}

\def\sigpar{\sigma_m}
\def\sigperp{\sigma_\theta}
\def\sigout{\sigma_z}
\def\sigin{\sigma_x}
\def\alphapt{\alpha^{\rm Pt}}
\def\alphapd{\alpha^{\rm Pd}}
\def\alphafe{\alpha^{\rm Fe}}
\def\uu{\upuparrows}
\def\ud{\uparrow\hspace{-0.055cm}\downarrow}
\def\psins{\psi^{\sigma}_n}
\def\psinsp{\psi_n^{\sigma^{\prime}}}
\def\psims{\psi_m^{\sigma}}
\def\psimsp{\psi_m^{\sigma^{\prime}}}
\def\psinso{\psi_{n,0}^{\sigma}}
\def\psinspo{\psi_{n,0}^{\sigma^{\prime}}}
\def\psimso{\psi_{m,0}^{\sigma}}
\def\psimspo{\psi_{m,0}^{\sigma^{\prime}}}
\def\psipso{\psi_{p,0}^{\sigma^{\prime\prime\prime}}}
\def\psipspo{\psi_{p,0}^{\sigma^{\prime\prime\prime}}}
\def\psilso{\psi_{l,0}^{\sigma^{\prime\prime}}}
\def\psilspo{\psi_{l,0}^{\sigma^{\prime}}}

\def\red#1{{\color{red}#1}}

\begin{abstract}

By performing first principles calculations we investigate the intrinsic anomalous Hall 
conductivity (AHC) and its anisotropy in ordered L1$_0$ FePt, CoPt and NiPt ferromagnets, and 
their intermediate alloys. We demonstrate that the AHC in this family of compounds depends strongly
on the direction of the magnetization $\mathbf{M}$ in the crystal. 
We predict that such pronounced orientational dependence in combination with the general 
decreasing trend of the AHC when going from FePt to NiPt leads to a sign change of the AHC
upon rotating the magnetization direction in the crystal of CoPt alloy. We also suggest that
for a range of concentration $x$ in Co$_x$Ni$_{1-x}$Pt and Fe$_x$Co$_{1-x}$Pt alloys it is possible to achieve a complete
quenching of the anomalous Hall current for a certain direction of the magnetization in the crystal.   
By analyzing the spin-resolved AHC in $3d$Pt alloys we endeavor to relate the overall trend of the
AHC in these compounds to the changes in their densities of $d$-states around the Fermi energy
upon varying the atomic number. Moreover, we show the generality of the phenomenon of anisotropic 
anomalous Hall effect by demonstrating its occurrence within the three-band tight-binding model.   

\end{abstract}

\maketitle

\section{Introduction}

Despite its long history, the anomalous Hall effect (AHE) in ferromagnets, 
discovered in 1881,\cite{Hall:1881} is still not fully understood from 
the theoretical point of view.\cite{Nagaosa:2010} 
Nevertheless, due to possible vast applications in spintronic devices, the AHE,
and its counterpart in nonmagnetic materials --- the spin Hall effect (SHE),
\cite{Hirsch:1999} have drawn quite intensive attention in the recent years. 
The underlying topological nature of the intrinsic AHE and SHE, relating
these phenomena to some fundamental physical effects, 
makes them even more relevant and interesting. 
Spin-orbit coupling (SOC) plays a crucial role in both AHE and SHE, 
as proposed in the first microscopic theory of the AHE
by Karplus and Luttinger. \cite{Karplus:1954} 
It can be demonstrated that SOC in perfect crystals 
gives rise to a transverse anomalous velocity of 
electrons propagating along the direction of the external electric field $-$ thus 
leading to the anomalous Hall current. 
This mechanism is nowadays referred to 
as the intrinsic contribution.

The intrinsic AHC considered in this work can be obtained via the linear response Kubo formula 
for the off-diagonal components of the conductivity tensor $\sigma$:
\begin{equation}
\label{eq:kubo}
\begin{split}
\sigma_{ij}\ &=-\ e^2\hbar\int_{\text{BZ}}\frac{d^3k}{8\pi^3}\,\Omega_{ij}(\mathbf{k}), \\
\Omega_{ij}(\mathbf{k})\ &=-\ 2{\rm Im} \sum_{n,m}^{o,e}\frac{\Braket{\psi_{n\mathbf{k}}|v_i|\psi_{m\mathbf{k}}}\Braket{\psi_{m\mathbf{k}}|v_j|\psi_{n\mathbf{k}}}}{(\varepsilon_{n\mathbf{k}}-\varepsilon_{m\mathbf{k}})^2},
\end{split}
\end{equation}
which relates the conductivity tensor to the Brillouin zone (BZ) integral of the $\mathbf{k}$-dependent
Berry curvature tensor $\Omega$. In the latter expression $\psi_{n\mathbf{k}}$ and 
$\psi_{m\mathbf{k}}$ are respectively the occupied ($o$) and empty ($e$) one-electron 
spinor Bloch eigenstates of the crystal, $\varepsilon_{n\mathbf{k}}$ and $\varepsilon_{m\mathbf{k}}$ 
are their eigenenergies, and $v_i$ and $v_j$ are the Cartesian component of the velocity operator
$\mathbf{v}$.
As a second-rank antisymmetric tensor, the AHC tensor can be also seen as the anomalous Hall
conductivity vector, $\boldsymbol{\sigma}$, the components of which 
are related to the components of the AHC as
$\sigma_{i}=\frac{1}{2}\epsilon_{ijk}\sigma_{jk}$, where $\epsilon_{ijk}$ 
is the Levi-Civita tensor. 

For materials with impurities or disorder, extrinsic contributions to the AHE also exist,
which can be described within density functional theory.\cite{Gradhand:2010,Lowitzer:2010PRL} 
Nevertheless, the 
intrinsic contribution is often dominating in itinerant ferromagnets with
moderate resistivity.\cite{Nagaosa:2010} 
Since the intrinsic anomalous Hall conductivity (AHC) is determined by the 
electronic structure of a pristine crystal (Eq. \ref{eq:kubo}), which can be accurately calculated
using modern first principles methods, a comparison between experiments 
and first principles calculations serves as the first necessary step to deeper 
understanding of the intrinsic AHE in real materials. Several investigations using the 
first principles methods have been done, for instance, in 
SrRuO$_3$,\cite{Fang:2003,Mathieu:2004} 
Fe,\cite{Yao:2004,Wang:2006} Mn$_5$Ge$_3$,\cite{Zeng:2006}
CuCr$_2$Se$_{4-x}$Br$_x$,\cite{Yao:2007} Ni,\cite{Wang:2007} 
Co.\cite{Wang:2007,Roman:2009}
For those materials, the calculated intrinsic AHC
agrees well with the experimental values, 
except for the case of fcc Ni,\cite{Wang:2007} which 
might be due to its complicated electronic structure. \cite{Yang:2001} 

One of the recently emerging topics in the field of the transverse magneto-transport phenomena is 
the anisotropic nature of the off-diagonal part of the conductivity tensor. \cite{Roman:2009, Freimuth:2010, Zhang:2011}
In the case of the AHE, the presence 
of the magnetization $\mathbf{M}$ in a ferromagnet leads to
a strong dependence of the components of the conductivity tensor on the magnetization direction
in the sample. Although experimentally, anisotropic AHE has been observed in many
materials,~e.g.~bcc Fe,\cite{Weissman:1973} fcc Ni,\cite{Volkenshtein:1961,Hiraoka:1968}
hcp Gd,\cite{Lee:1967} as well as FeCr$_2$S$_4$,\cite{Ohgushi:2006}
Yb$_{14}$MnSb$_{11}$, \cite{Sales:2008} Y$_2$Fe$_{17-x}$Co$_x$\cite{Skokov:2008}
and $R_2$Fe$_{17}$ ($R$ = Y, Tb, Gd),\cite{Stankiewicz:2011}
only two studies of the anisotropy of the AHE from first principles have been performed so far.
Roman {\it et al.}\cite{Roman:2009} considered the anisotropic AHE in uniaxial hcp Co, calculated 
the ratio of the AHCs for the out-of-plane and in-plane magnetization,
$\sigma_z$ and $\sigma_x$, respectively, and found it to be as large as four, which is close to the experimentally observed 
ratio.\cite{Volkenshtein:1961} Moreover, they performed a directional averaging of the anisotropic AHC 
and compared the obtained conductivity to the experimental value measured in polycrystalline hcp Co samples,\cite{Gil:2005} finding an excellent agreement.\cite{Roman:2009} In another work, Zhang and 
co-authors \cite{Zhang:2011} considered the anisotropic AHE in uniaxial L1$_0$ FePt alloy.
They also found a large anisotropy of the AHC in this compound, and were able to attribute it
to the spin-non-conserving part of the spin-orbit interaction, prominent in this material with
strong SOC. 

In this work, we undertake a detailed first principles analysis of the anisotropic intrinsic AHE in the 
group of L1$_0$-ordered $3d$Pt ($3d$ = Fe, Co, Ni) alloys. These materials are currently under 
investigation with respect to possible spintronic applications due to their large uniaxial magnetic
anisotropy energies and high Kerr rotation, making them possible candidates for ultrahigh density 
magnetic and magneto-optical recording media. \cite{Cebollada:1994}
Recently, the AHE in FePt was used for injection of a spin-polarized current for consequent detection 
of direct and inverse spin Hall effect in Au. \cite{Seki:2008}
In a combined experimental and theoretical study, \cite{Seemann:2010} it was shown that the intrinsic contribution
to the anomalous Hall signal dominates in samples of FePt with finite structural disorder. 
All this motivated our study of the anisotropy of the intrinsic AHE in uniaxial
$3d$Pt alloys. 

In general, we find very large 
anisotropy of the AHE in these compounds, which changes its magnitude and sign as a function
of the band filling 
of the $3d$ transition-metal.
In particular, we observe that for the $3d$Pt alloys with high concentration
of Co atoms the $\sigout$ and $\sigin$ conductivities differ in sign, which leads to the phenomenon
of the {\it anti-ordinary} AHE, in which at a certain "magic" angle of the magnetization
the Hall current $\mathbf{J}$ becomes parallel to $\mathbf{M}$. Moreover, for 
(Fe$_{0.1}$Co$_{0.9}$)Pt and (Co$_{0.85}$Ni$_{0.15}$)Pt alloys we predict the occurrence of the
{\it colossal} anisotropy of the AHE, that is, an order of magnitude reduction in the value of
$\sigin$ as compared to $\sigout$, or visa versa. By analyzing the spin-resolved AHC in these
alloys, we try to relate the general trend of decreasing AHC in these compounds when going from 
FePt to NiPt to the changes in their densities of states around the Fermi energy. 
Moreover, we demonstrate the occurrence of the anisotropic AHE within the "uniaxial" minimal 
three-band tight-binding model, which underlines the generality of this phenomenon and hints at 
its occurrence in a wide range of materials. 

The structure of the paper is as follows. In Section~II we describe the method and details of 
our first principles calculations. In Section~III we introduce a minimal three-band $t_{2g}$
model, necessary to predict the appearance of the anisotropy of the AHE in a three-dimensional
crystal, and investigate the AHC within this model as a function of the band filling. In Section~IV
we present the results of our {\it ab initio} calculations of the AHE in the family of ordered FePt, CoPt and
NiPt alloys. We demonstrate that the AHE in these alloys in strongly anisotropic and displays 
a number of interesting phenomena in the region where it changes sign. We end the paper with conclusions.

\section{Computational details}

We performed our density-functional theory (DFT) calculations of L1$_0$ ordered 
$3d$Pt ($3d$ = Fe, Co, Ni) alloys using the full-potential linearized augmented 
plane-wave (FLAPW) method as implemented in the J\"ulich DFT code {\tt FLEUR}.\cite{fleur} 
The generalized gradient approximation (GGA)\cite{Perdew:1996} for the 
exchange-correlation potential was used. The self-consistent calculations with SOC 
were done in second variation with k$_{\rm max}$ of 4.0~a.u.$^{-1}$ and 16000 $k$-points 
in the full Brillouin zone (BZ). The muffin-tin radius of 2.45~a.u. was used for all atoms. 
Six local orbitals for the 4$p$-states of Pt atoms were used to take care of the core charge of Pt 
correctly. In all our calculations, a $tp_2$ geometry 
with two atoms in the L1$_0$ phase was used for all alloys, with experimental lattice 
constants (Fig.~1).\cite{Ravindran:2001}
For intermediate alloys, for instance, (Fe$_{0.5}$Co$_{0.5}$)Pt, the virtual crystal
approximation (VCA) was applied on the $3d$ atomic sites, where the composition-averaged core
potential is used instead of that of pure $3d$ elements, together with corresponding 
number of valence electrons, and interpolated lattice constants
from the neighboring compounds. 

\begin{figure}
\includegraphics[width=5.5cm]{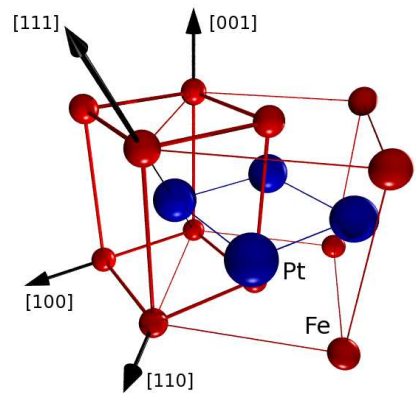}
\caption{(color online) \label{fig:struct}
Crystal structure of L1$_0$ FePt alloy. Small (red) spheres stand for the Fe atoms, while
large (blue) spheres mark the Pt ions. The primitive unit cell used in the calculations is
enclosed with thicker lines. In the text, $z$ stands for the [001] axis, while $x$ stands
for the [110] direction in the crystal.
}
\end{figure}

In this work, we applied the Wannier interpolation technique to calculate the AHC accurately.
We followed the method introduced in Refs.~[\onlinecite{Souza:2002}] and [\onlinecite{Freimuth:2008}]
to construct the maximally-localized Wannier functions (MLWFs) from the FLAPW wave functions, 
in which the unitary transformations are constructed to minimize the spread of the Wannier functions.
Using the self-consistent charge density with SOC included, 36 MLWFs corresponding to
$s,p,d$-orbitals of $3d$ and Pt atoms for both spins were generated 
on a $10\times10\times10$ $k$-mesh, 
using the {\tt WANNIER90} code.\cite{wannier90}
The AHC was then calculated by applying the Wannier interpolation 
scheme of Wang {\it et al.}\cite{Wang:2006} for evaluating the Berry curvature  
on a $208\times208\times208$ uniform $k$-mesh. For $k$-points at which the Berry 
curvature exceeded 30 \AA$^2$, an adaptive refined $k$-mesh of 
$5\times5\times5$ was used. 

\section{AHE anisotropy: Generalities}

In terms of the AHC vector the linear response expression for the 
anomalous Hall current $\mathbf{J}$ can be rewritten as 
\begin{equation}
\mathbf{J}(\mathbf{M})=\boldsymbol{\sigma}(\mathbf{M})\times\mathbf{E},
\end{equation}
where $\mathbf{E}$ is the electric field.
 In a ferromagnet with uniform magnetization $\mathbf{M}$, $\boldsymbol{\sigma}$ and $\mathbf{J}$ depend 
 on the magnetization direction in the crystal. This magnetocrystalline anisotropy of the AHC manifests itself in the changes in the direction and magnitude of $\boldsymbol{\sigma}$ upon changing the direction of the magnetization away from a certain (easy) axis.
In general, while $\mathbf{J}$ 
is always perpendicular to the electric field $\mathbf{E}$, it is not necessarily orthogonal
 to $\mathbf{M}$, as $\boldsymbol{\sigma}$ and $\mathbf{M}$ may not
be parallel. In single
crystals ${\boldsymbol\sigma}$ and $\mathbf{M}$ are perfectly collinear only when
$\mathbf{M}$ points along certain high symmetry directions. For an arbitrary
orientation of $\mathbf{M}$ there is generally a misalignment between them, which
is the signature of the anisotropic AHE. Another manifestation of the
AHE anisotropy is the dependence of the absolute value of $\mathbf{J}$ and
${\boldsymbol\sigma}$ on the direction of $\mathbf{M}$. While in cubic
crystals the AHC anisotropy appears only at the 3rd order with respect to the directional
cosines of the magnetization, in uniaxial crystals the linear term can dominate,\cite{Roman:2009} 
leading to large AHE anisotropies, observed experimentally\cite{Volkenshtein:1961} and explained
theoretically.\cite{Roman:2009} 

The microscopic origin of the anisotropic AHE can be easily understood by inspecting Eq.~(\ref{eq:kubo}).
Consider a tetragonal crystal structure, as depicted for L1$_0$ FePt alloy in Fig.~1. Suppose that we start with the magnetization $\mathbf{M}$ pointing along the [001] axis ($z$-axis
in the following). In this case the $v_x$ and $v_y$ components of the velocity operator have to be
inserted at the place of $v_i$ and $v_j$ operators in Eq.~(\ref{eq:kubo}) in order to obtain the
$\sigma_{xy}$ component of the conductivity tensor, or, equivalently, the $\sigma_z$ component
of the conductivity vector, taking into account that for such a high symmetry direction of the
magnetization $\boldsymbol{\sigma}$ is aligned with $\mathbf{M}$ along the $z$-axis. 
Rotating now $\mathbf{M}$ away from the [001] axis 
modifies (i) the wave functions $\psi_{n\mathbf{k}}$ and $\psi_{m\mathbf{k}}$, (ii) occupation
of the states and (iii) the eigenenergies of the states $\varepsilon_{n\mathbf{k}}$ and 
$\varepsilon_{m\mathbf{k}}$ due to the presence of the spin-orbit interaction. 
Thus,
all components of the conductivity tensor have to be recalculated
for a general direction of $\mathbf{M}$. 
In this work, the magnetization is confined in the high-symmetry
($\bar{1}$10)-plane, and the resulting conductivity vector also lies in the same plane due to
the antisymmetric nature of the anomalous Hall conductivity with respect to the inversion of the 
magnetization direction.
For a general magnetization direction $\mathbf{M}$ in this plane, 
the AHC vector $\boldsymbol{\sigma}$ can be decomposed as follows:
\begin{equation}
\mathbf{\boldsymbol\sigma}=\mathbf{\sigma_{\parallel}}\hat{\mathbf{M}} +\mathbf{\sigma_{\perp}}\mathbf{n},
\end{equation}
where $\hat{\mathbf{M}}$ and $\mathbf{n}$ are the unit vectors along the magnetization
direction and orthogonal to it within the ($\bar{1}$10)-plane, respectively. The ratio of $\sigma_{\Vert}(\mathbf{M})$
and $\sigma_{\perp}(\mathbf{M})$ gives an estimate of how strongly the AHC vector deviates from the direction
of $\mathbf{M}$. Upon further rotation the magnetization hits the [110] direction in the crystal ($x$-axis
in the following),  and the orthogonal component of the AHC, $\sigma_{\perp}$, is zero, while
$\boldsymbol{\sigma}$ is collinear with the magnetization again. In this case $v_y$ and $v_z$
enter Eq.~(\ref{eq:kubo}), and the magnitude
of the AHC is given by $\sigin$.

\section{Anisotropic AHE within the $t_{2g}$-model}

In this section we demonstrate the appearance of anisotropic AHE within a simple tight-binding model,
namely, three-band $t_{2g}$ model for $d_{yz}$, $d_{zx}$, and $d_{xy}$ spin-up 
orbitals on a cubic lattice. We consider only the hoppings up to the nearest neighbors, $t_1$, and
to the next nearest neighbors, $t_2$. The Hamiltonian of the model in $k$-space reads: 
\begin{equation}
\label{eq:t2g}
H(\mathbf{k})=H_0(\mathbf{k})+H_{\text{SO}}(\mathbf{M}),
\end{equation}
where the Hamiltonian without SOC is given by:
\begin{widetext}
\begin{equation}
H_0(\mathbf{k})=\left(\begin{array}{ccc} 
            -2t_1(\cos{k_y}+\text{A}\cos{k_z})   &      4t_2\sin{k_x}\sin{k_y}     &     4t_2\sin{k_x}\sin{k_z}  \\
             4t_2\sin{k_x}\sin{k_y}      & -2t_1(\cos{k_x}+\text{A}\cos{k_z})      &     4t_2\sin{k_y}\sin{k_z}  \\ 
             4t_2\sin{k_x}\sin{k_z}      &      4t_2\sin{k_y}\sin{k_z}     &  -2t_1(\cos{k_x}+\cos{k_y}) \end{array} \right),
\end{equation}
\end{widetext}
in which we introduced an anisotropy parameter $A$. The role of this parameter is
to make the system uniaxial,~i.e., for $A\neq 1$ the nearest neighbor hopping in the $(x,y)$-plane is different from that along the $z$-axis. In a real cubic crystal introducing such a uniaxiality could correspond to~e.g.~changing the interlayer distance along the $z$-axis via application of stress.

\begin{figure}[t!]
\includegraphics[width=8.4cm]{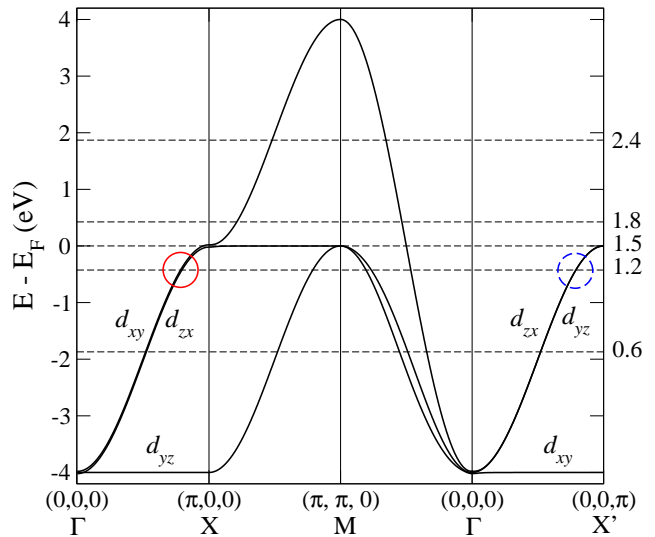}
\caption{
Electronic bands of the $t_{2g}$-model for $\mathbf{M}\Vert x$, with
$t_2/t_1=0.1$,  $\xi/t_1=0.02$, $A=1.0$, and the lattice constant
of 1~\AA. The dashed horizontal lines mark the position of the 
Fermi level for the electronic occupation given by the number on the 
right. Labels mark the orbital character of the bands. \label{fig:t2g-bands}}
\end{figure}

The $k$-independent SOC part of the Hamiltonian depends on the magnetization
direction $\mathbf{M}$ in the crystal. Within our model for $\mathbf{M}$ along 
the $z$-axis the $d_{yz}^{\uparrow}$- and $d_{zx}^{\uparrow}$-orbitals are 
coupled due to SOC, and $H_{\rm SO}$-matrix reads:
\begin{equation}
H_{\rm SO} ( \mathbf{M}\Vert z) =\xi\left(\begin{array}{ccc}
      0 & i & 0 \\ 
     -i & 0 & 0 \\
      0 & 0 & 0 \end{array}\right),
\end{equation}  
while for $\mathbf{M}$ along the $x$-axis $d_{zx}^{\uparrow}$- and $d_{xy}^{\uparrow}$-orbitals are coupled instead:
\begin{equation}
H_{\rm SO} (\mathbf{M}\Vert x) = \xi\left(\begin{array}{ccc}
      0 & 0 & 0 \\
      0 & 0 & i \\ 
      0 &-i & 0 \end{array}\right).
\end{equation}
The strength of the spin-orbit interaction is constant in both cases and is given by
parameter of the model $\xi$. The band structure of the model obtained by diagonalizing Hamiltonian~(\ref{eq:t2g})
is plotted in Fig.~\ref{fig:t2g-bands} for $\mathbf{M}\Vert x$ and $A=1$.

\begin{figure}[]
\includegraphics[width=8.6cm]{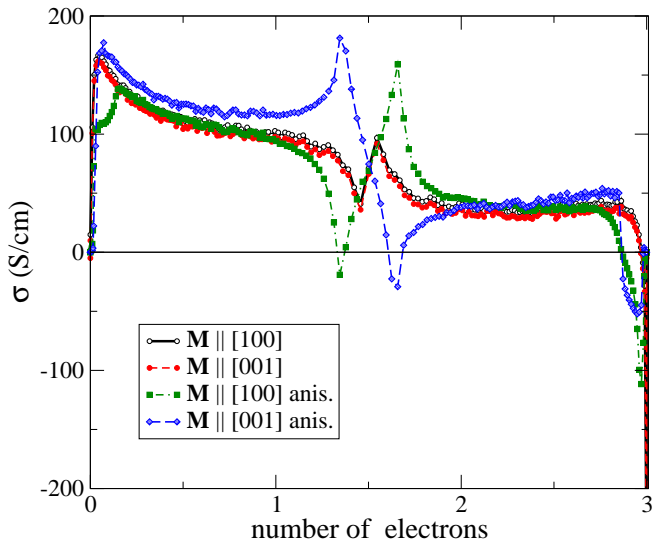}
\caption{Anomalous Hall conductivity as a function of the 
band filling within the three-band $t_{2g}$ model with the
parameters given in the caption to Fig.~\ref{fig:t2g-bands}.
Open (filled) circles mark the case of 
$\mathbf{M}\Vert x(z)$ in the "isotropic" crystal with $A=1$ in 
Eq.~(\ref{eq:t2g}).
Note that in this case the AHC curve for $\mathbf{M}\Vert z$
is shifted by 5~S/cm with respect to the AHC with $\mathbf{M}\Vert x$
in order to see the degeneracy between the two curves clearly.
Filled squares (diamonds) stand for the AHC in the "anisotropic"
crystal with $A=0.9$ in Eq.~(\ref{eq:t2g}) for $\mathbf{M}\Vert x(z)$. 
\label{fig:t2g}}
\end{figure}

The AHC calculated according to Eq.~(\ref{eq:kubo}) for the $t_{2g}$-model is 
shown in Fig.~\ref{fig:t2g} as a function of band-filling for $\mathbf{M}\Vert z$
and $\mathbf{M}\Vert x$, both with $A=1$ and $A=0.9$. We assumed for our
calculations a lattice constant of 1~\AA, $t_2/t_1=0.1$ and $\xi/t_1=0.02$.
Obviously, as expected from symmetry, the AHC does not depend on whether 
the magnetization points along the $x$- or $z$-axis when $A=1$, while it displays 
a strong dependence on the electron occupation $n$, or, equivalently, on the position 
of the Fermi level
$E_F^{n}$ which corresponds to this occupation. Such sensitive dependence of
intrinsic AHC on the details of the electronic structure, which stems from a very 
irregular behavior of the Berry curvature in the Brillouin zone, is rather 
well-known.\cite{Nagaosa:2010}
From our calculations it can be seen that when the Fermi energy is positioned 
in the vicinity of the band edges with high density of electronic states (DOS) $-$ which in our model corresponds to the case 
of nearly filled, half-filled and completely filled bands $-$ the
$E_F$ position has a strong effect on the AHC (Fig. \ref{fig:t2g}).
The reason behind such a sensitive dependence lies in the presence of flat
degenerate bands around the Fermi energy, which provide wide regions in
$k$-space where the occupied and unoccupied states are separated by a 
small energy.
We speculate that such a situation can also be related to the anomaly of the
density of states  near the band edges and associated Lifshitz transitions.\cite{Hongbin:phd} 
We expect that at such transitions, anomalies in the 
AHC would lead to~e.g.~singular behavior of the anomalous thermopower.
  
In case of the "uniaxial" $t_{2g}$-model ($A=0.9$) the electronic structure
of the crystal with $\mathbf{M}\Vert x$ and $\mathbf{M}\Vert z$ is not the same
anymore, and a strong anisotropy of the AHC can be seen in Fig.~\ref{fig:t2g},
leading even to a difference in sign, considered in detail in the
following section. In analogy to the case of the "isotropic" crystal, the largest 
difference between the AHC for the two magnetization directions can be observed 
for $n\approx 0$, $n\approx 1.5$ and $n\approx 3$, although for exact 
half-occupation ($n = 1.5$), this difference vanishes. 

In order to see the origin of the anisotropic AHC in the uniaxial
$t_{2g}$-crystal clearly, we refer to the band structure of the system, 
Fig.~\ref{fig:t2g-bands}, for $\mathbf{M}\Vert x$ and $A=1$. 
Let us consider the case of $n=1.2$ and transitions between the occupied and 
unoccupied states in the energy region marked with red solid circle along the
$\Gamma X$-path. For $\mathbf{M}\Vert x$ the SOC leads to the mixing of $d_{xy}$-
and $d_{zx}$-orbitals, and the resulting small energy splitting between the two
corresponding bands can be clearly seen. 
The non-vanishing matrix element of the SOC between the two latter orbitals
leads to a finite contribution to the Berry curvature and the AHC, associated with 
the electronic transitions across $E_F^{1.2}$. On the other hand, the $d_{zx}$-
and $d_{yz}$-orbitals are not coupled by SOC for this magnetization direction 
(which can be also seen from an exact degeneracy of corresponding bands along $\Gamma X'$),
and the contribution from the symmetry-equivalent without SOC part of the band structure
along $\Gamma X'$ to the Berry curvature, marked with dashed blue circle, is 
exactly zero. The situation is reversed for $\mathbf{M}\Vert z$ in the
isotropic crystal, and the contribution to the AHC from the states in the dashed 
blue circle is exactly the same as that from the corresponding region along the 
$\Gamma X$-path for $\mathbf{M}\Vert x$, while the latter gives no contribution
for $\mathbf{M}\Vert z$. Overall, when only the encircled regions and their
symmetric "clones" are considered, the resulting AHC does not depend on the 
magnetization direction and there is no anisotropy of it. Introducing now anisotropy
in the system by setting $A$ to $0.9$ in the $t_{2g}$-model leads to the fact
that the electronic structure along the $\Gamma X$- and $\Gamma X'$-paths is
not the same anymore and thus, the contributions to the AHC from the full red
circle for $\mathbf{M}\Vert x$ and from the dashed blue circle for 
$\mathbf{M}\Vert x$ are different. In turn, this leads to the AHE anisotropy.
This line of thinking is clearly valid also for explaining the anisotropy of the
total AHC, which is obtained by a summation over all such encircled
regions in energy and $k$-space, contributing to the AHC.  

\section{Anisotropic AHC in  $3d{\rm Pt}$ alloys}

\begin{figure}
\centering
\includegraphics[width=8.5cm]{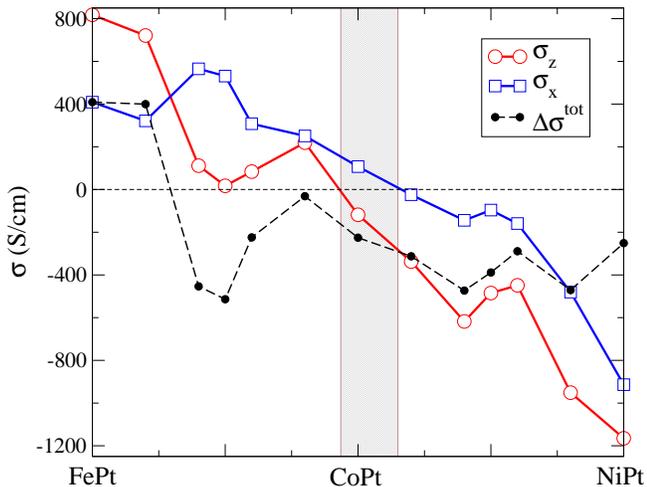}
\caption{
Anomalous Hall conductivity of $3d$Pt alloys
for $\mathbf{M}$ along [001] ($\sigout$, open circles) and [110] ($\sigin$,
open squares), and anisotropy ($\Delta\sigma^{tot}=\sigout - \sigin$, filled circles)
, with respect to the band filling. 
\label{fig1}}
\end{figure}

The results of our calculations for the intrinsic AHC in L1$_0$-ordered FePt, CoPt and NiPt 
and their intermediate alloys are presented in Table~\ref{table:aahe-3dpt} and in Fig.~\ref{fig1} for [001] ($\sigout$) and [110] ($\sigin$) directions of the 
magnetization $\mathbf{M}$ in the crystal. In general, the observed behavior of $\sigout$
and $\sigin$ as a function of the electron occupation of the $3d$ transition-metal is similar: starting from the FePt
alloy with positive values of the AHC for both magnetization directions of around 600~S/cm, the sign of $\sigout$
and $\sigin$ changes in the vicinity of the CoPt alloy, and the AHC values are very large
and negative for NiPt, reaching as much as $-1200$~S/cm. On average, we can see 
that the decrease of both conductivities with increasing electron occupation is rather 
linear. It is worth mentioning that such decreasing trend of the AHC is somewhat 
reminiscent of the trend among the pure bcc Fe, hcp Co and fcc Ni, for which the
calculated intrinsic values of the $\sigout$ AHC constitute approximately 750~S/cm,
\cite{Yao:2004,Wang:2006} 480~S/cm,\cite{Wang:2007,Roman:2009} and $-$2200~S/cm,
\cite{Wang:2007} respectively. We will come back to this point at the end of this section. 

\begin{table}
\begin{ruledtabular}
\begin{tabular}{lr|rrrrrr}
    &   & $\sigma^{\rm tot}$ & $\sigma^{\uu}$ & $\sigma^{\ud}$ & $\Delta\sigma^{\rm tot}$ & $\Delta\sigma^{\uu}$ & $\Delta\sigma^{\ud}$ \\ \hline
FePt & $\sigout$ & 818       &     577     &      133  &  409     &    $-$9   &     317   \\
       & $\sigin$   & 409       &     585     &  $-$184 &            &              &             \\ \hline
CoPt & $\sigout$ & $-$119  &      487    &  $-$513  & $-$226 &  $-$7     &  $-$210 \\
       & $\sigin$   & 107       &      494    &  $-$303  &            &              &             \\ \hline
NiPt & $\sigout$ & $-$1165 &  $-$1495 &  $-$550  & $-$251 &  $-$1215 & 7         \\
       & $\sigin$   & $-$914  &  $-$280    &  $-$557 &            &               &            \\
\end{tabular}
\end{ruledtabular}
\label{table:anisotropy}
\caption{
  Values of the AHC in L1$_0$ FePt, CoPt and NiPt with the magnetization along
  [001] ($\sigout$) and [110] ($\sigin$). For each orientation,
  $\sigma^\uu$ ($\sigma^{\ud}$) is calculated by keeping only the
  first (second) term in the spin-orbit Hamiltonian (Eq. \ref{eq:soi}), while both terms
  are kept when calculating $\sigma^{\rm tot}$.  
  $\Delta\sigma^{\rm tot}$ is defined as $\sigout-\sigin$.
  $\Delta\sigma^{\uu(\ud)}$ is defined as the difference between the
  spin-conserving (spin-flip) parts of $\sigma_z$ and $\sigma_x$.
  All values are in S/cm.\label{table:aahe-3dpt}}
\end{table}

It is clear from Table~\ref{table:aahe-3dpt} and Fig.~\ref{fig1} that
for almost all considered 
alloys the anisotropy of the AHC reaches very large values. This is expected,
since in uniaxial crystals the AHC anisotropy appears already in the first order with
respect to the directional cosines of the magnetization (see discussion in the previous section). \cite{Roman:2009} 
In FePt, the difference between $\sigout$ and $\sigin$, $\Delta\sigma^{\rm tot}$ (filled circles in Fig. \ref{fig1}),
is as large as $\sigin$ itself and constitutes around 400~S/cm,~c.f.~Table~\ref{table:aahe-3dpt}.\cite{Zhang:2011} In CoPt, on the other hand, 
the absolute value of $\Delta\sigma^{\rm tot}$ is twice larger
than the absolute value of the AHC for any of the two magnetization directions. 
The AHC anisotropy  reaches as much as $-$500~S/cm in the vicinity of FeCoPt
and CoNiPt alloys, and in general, the behavior of $\Delta\sigma^{\rm tot}$ is neither smooth
nor monotonous, displays several mimina and maxima as a function of the electron 
occupation, and even changes its sign for Fe$_x$Co$_{1-x}$Pt alloy with $x \approx 0.75$.
On the other hand, the anisotropy of the AHE in $3d$Pt alloys, when the magnetization
is rotated in the (001)-plane, is much smaller than the "out-of-plane$-$in-plane" anisotropy discussed previously, 
which can be easily understood taking into consideration the higher symmetry of the former
situation. E.g., the difference of the AHCs for $\mathbf{M}$
along [110] and [100] reaches at most 80~S/cm in NiPt alloy, being as small as
$-$47~S/cm in CoPt and $-$16~S/cm in FePt. 

With grey shaded area in Fig.~\ref{fig1} we highlight the region around the
CoPt alloy, where both $\sigout$ and $\sigin$ change their sign. This sign change leads 
to the occurrence of two interesting phenomena with respect to the anisotropic AHE. 
The first one, which we name the {\it colossal anisotropy} of the AHE, according to 
our calculations, occurs for Fe$_x$Co$_{1-x}$Pt alloy with $x\approx 0.1$ and for  Co$_x$Ni$_{1-x}$Pt alloy with $x\approx 0.85$. For these two compounds one of the conductivities,
$\sigout$ for Fe$_{0.1}$Co$_{0.9}$Pt and $\sigin$ for Co$_{0.85}$Ni$_{0.15}$Pt,
turns to zero, which marks the complete disappearance of the intrinsic anomalous Hall 
current $\mathbf{J}$ for one of the magnetization directions in the crystal. We introduce
the term colossal anisotropy in analogy to the situation which was predicted to occur in 
one-dimensional Pt wires, for which upon changing the magnetization direction $\mathbf{M}$ the value of the magnetization $|\mathbf{M}|$ itself can be quenched completely.\cite{Smogunov:2008} In terms of the longitudinal transport within the setup of~e.g.~anisotropic magnetoresistance (AMR) experiment, the occurrence of the colossal anisotropy of the diagonal conductivity would results in a metal-insulator transition in the crystal $-$ in case of the 
colossal AHE anisotropy observed in $3d$Pt alloys all compounds remain metallic for all
magnetization directions, however, and retain their complicated electronic 
structure around the Fermi energy.

\begin{figure}
\centering
\includegraphics[width=8.3cm]{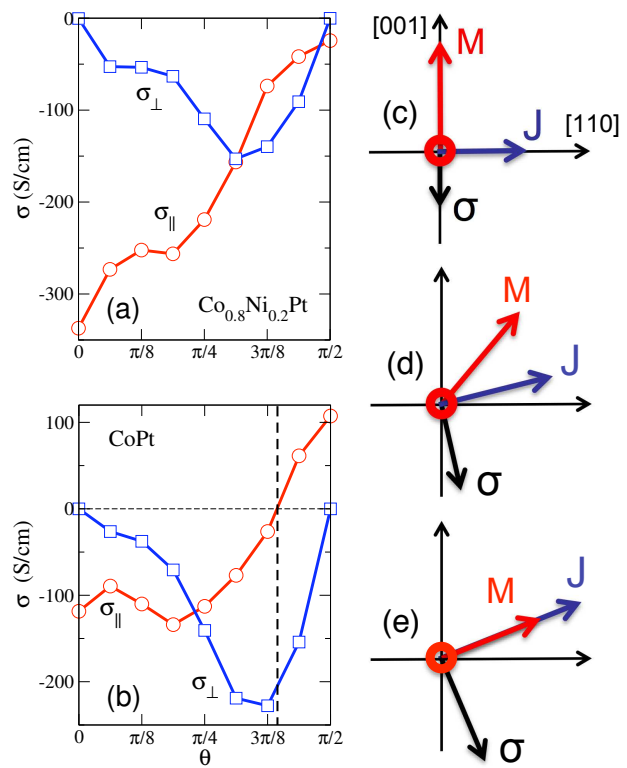}
\caption{(a) Colossal anisotropy of the AHC in Co$_{0.8}$Ni$_{0.2}$Pt alloy.
Red circles (blue squares) denote the $\sigma_{\parallel}$
($\sigma_{\perp}$) component of the AHC as a function of the angle 
$\theta$ of the magnetization $\mathbf{M}$ with [001]-axis upon rotating it into the 
[110] direction. (b) Anti-ordinary Hall effect in CoPt. 
Red circles (blue squares) denote the $\sigma_{\parallel}$ 
($\sigma_{\perp}$) component of AHC, as a function of the angle 
$\theta$ of the magnetization $\mathbf{M}$ with [001]-axis upon 
rotating it into the [110] direction. (c)-(e) depict the relative orientation of the Hall current $\mathbf{J}$,
AHC $\boldsymbol{\sigma}$ and magnetization $\mathbf{M}$ in the situation of the
anti-ordinary AHE. In (c)-(e) the magnetization is confined to the ($\bar{1}$10)-plane.
\label{fig2}}
\end{figure}

For Co$_{0.8}$Ni$_{0.2}$Pt alloy in Fig.~\ref{fig2}(a) we plot the dependence of the $\sigma_{\Vert}$ 
and $\sigma_{\perp}$ components of the AHC on the angle $\theta$ of the 
magnetization with the $z$ axis when it is  rotated away from the 
[001] direction within the ($\bar{1}$10) plane 
towards the [110] direction. At $\theta=0$ the $\sigma_{\perp}$ component is 
zero and the AHC vector with magnitude of 340~S/cm is antiparallel to the $z$ axis, along which
the magnetization is aligned,~c.f.~Fig.~\ref{fig2}(c). 
Upon increasing $\theta$ we observe the increase in $\sigma_{\Vert}$ and 
decrease in $\sigma_{\perp}$, with both components becoming equal at the angle $\theta\approx 55^{\circ}$. 
At this angle the magnitude of the AHC is reduced significantly to
210~S/cm, while its deviation from the $z$-axis is only about 10$^{\circ}$. Thus, in this
range of $\theta$, the rotation of the magnetization results mainly in quenching the magnitude
of the anomalous Hall current, while its direction basically remains "stuck" to the [110] axis. Upon further
rotation of the magnetization both components of the AHC vector rapidly approach zero, the AHC vector
quickly rotates  towards the $-x$ axis, and when $\mathbf{M}$ hits the [110] direction, the AHC
with a tiny magnitude of 25~S/cm is again antiparallel to the magnetization.    
 
For CoPt alloy the situation, depicted in Fig.~\ref{fig2}(b-e), is completely different. Similarly to the
previously considered case, at $\theta=0^{\circ}$ the AHC vector is antiparallel to $\mathbf{M}$, and 
its magnitude constitutes 120~S/cm, Fig.~\ref{fig2}(c). Upon increasing $\theta$ up to as much as 45$^{\circ}$
the conductivity vector resides basically in the close vicinity of the [00$\bar{1}$] axis, while its magnitude 
increases. For example at $\theta=45^{\circ}$, $\sigma_{\Vert}\approx\sigma_{\perp}$ and the value
of total $\sigma$ is roughly 170~S/cm, Fig.~\ref{fig2}(d). With further increasing $\theta$ the magnitude
of the AHC is increasing even further, while the AHC vector starts its way towards the [110]-direction. The increase
of $|\sigma|$ is mainly due to the $\sigma_{\perp}$ component in this regime, while at the same time 
$|\sigma_{\Vert}|$ is becoming smaller, and eventually changes its sign. Finally, at $\theta=90^{\circ}$,
the AHC vector is aligned together with $\mathbf{M}$ along the $x$ axis, and its magnitude is 110~S/cm.

Remarkably, $\sigma_{\Vert}$ turns to zero at $\theta_0=70^{\circ}$, which manifests the occurrence of the  {\it anti-ordinary} Hall effect in the crystal of CoPt, see Fig.~\ref{fig2}(e). At this "magic" angle, the magnitude of the anomalous Hall current $\mathbf{J}$ is almost twice larger than it is for $\mathbf{M}\Vert z$, however,
due to non-vanishing $\sigma_{\perp}$ component of the AHC vector, $\mathbf{J}$ is aligned {\it along} the 
direction of the magnetization. By analyzing Figs.~\ref{fig2}(b-e) we observe that the rotational sense
of the anomalous Hall current is opposite to that observed in the ordinary Hall effect (OHE) of free electron gas. 
For OHE, Lorentz forces $\sim[\mathbf{H}\times\mathbf{v}]$ are acting on electrons with velocity $\mathbf{v}$ 
in the presence of magnetic field $\mathbf{H}$.
The resulted ordinary Hall current of free electrons is always perpendicular to $\mathbf{H}$ irrespective of its
direction, opposite
to the situation of the anti-ordinary anomalous Hall effect, observed in CoPt. Here, turning the magnetization clockwise 
in the $(\bar{1}10)$-plane results in an anti-clockwise rotation of $\mathbf{J}$, with its value staying rather large
all the time. The anti-ordinary {\it spin} Hall effect has been also recently predicted to occur in 
transition metals. \cite{Freimuth:2010}

In the region of $3d$Pt alloys in the vicinity of L1$_0$ CoPt the anisotropy of the AHE manifests itself in crucial 
ways suggesting new functionalities of the AHE-based
devices. In this region, large changes in the magnitude of the anomalous Hall current as well as relative orientation
of the Hall current with respect to the magnetization can be easily achieved by simple reorientation of the sample's magnetization. While the former could be used in order to~e.g.~tune the relative magnitudes of the extrinsic and
intrinsic anomalous Hall signal,\cite{Seemann:2010, Lowitzer:2010PRL} 
among most straightforward applications of the latter could be a realization of the 
planar Hall effect (PHE),\cite{Yau:1971} which is related to the Hall effect in ferromagnetic materials observed in a two-dimensional
geometry with electric field, magnetization and the Hall current sharing same sample plane. So far, it is believed
that in most of the cases the PHE originates from anisotropic magnetoresistance in metallic ferromagnets, 
although the PHE mechanism stemming from the anomalous Hall effect due to non-collinearity of the magnetization in semiconductor-based materials
has been also suggested.\cite{Bowen:2005} Within the scope of the anti-ordinary Hall effect, described in this work, it would be possible 
to observe the PHE coming solely from the anisotropic nature of the collinear ferromagnetic materials.
 
We would like to underline, that despite the crudeness of the VCA approximation for description of the
electronic structure of complex alloys, the results of our work still hold, although the exact width of the
region where the colossal anisotropy and anti-ordinary nature of the intrinsic anomalous Hall effect can be
observed, might be different when more appropriate approximations, such as coherent potential approximation (CPA),\cite{Lowitzer:2010PRL} 
are used to treat the substitutional alloys Fe$_x$Co$_{1-x}$Pt and Co$_x$Ni$_{1-x}$Pt. The main reason behind this is that for "pure"
ferromagnets FePt, CoPt and NiPt
our results are exact in the sense that no approximations of disorder need to be made
and the AHE consists only of the intrinsic contribution, while the precise value of the intrinsic
AHC will still depend on the chosen parameters and formulations of the DFT calculations such 
as exchange-correlation functionals, treatment of SOC, validity of the single-particle picture,
particular choice of the basis set, {\it etc.}
The values of the intrinsic AHC at the ends of the considered family of alloys,
namely, FePt and NiPt, are large in their magnitude but differ in their sign. This means that upon varying
the concentration $x$ in Fe$_x$Co$_{1-x}$Pt and Co$_x$Ni$_{1-x}$Pt alloys, the region where the AHC changes
sign must exist, irrespective of the approximations made. At the end, it is the change of sign of the AHC 
for the CoPt alloy which leads to the occurrence of the colossal anisotropy and anti-ordinary anomalous 
Hall effect in its vicinity according to our calculations. 
 
At the end of this section we will try to relate the mentioned above change of sign in the values of the AHC
when going from FePt to NiPt, to the changes in the electronic structure of these materials. For this
purpose, we first of all decompose the atomic spin-orbit Hamiltonian in the well-known way: \cite{Zhang:2011}
  \begin{equation}
  \label{eq:soi}
    \xi\mathbf{L\cdot S}=
    \xi{\rm L}_{\hat{n}}{\rm S}_{\hat{n}} + \xi\left( 
    {\rm L}^{+}_{\hat{n}}{\rm S}^{-}_{\hat{n}} + 
    {\rm L}^{-}_{\hat{n}}{\rm S}^{+}_{\hat{n}} \right)/2,
  \end{equation}
where $\xi$ is the spin-orbit coupling strength, $\hat{n}$ is the spin magnetization direction 
(which is taken as the spin-quantization axis), $\mathbf{L}$ and $\mathbf{S}$ are the total
orbital and spin angular momentum operators, ${\rm L}_{\hat{n}}=\mathbf{L}\cdot\hat{n}$, 
and ${\rm L}^{+}_{\hat{n}}$ and ${\rm L}^{-}_{\hat{n}}$ are the corresponding raising and 
lowering operators (analogously for spin). We shall refer to the first and second terms in Eq.~(\ref{eq:soi}) 
as the spin-conserving  and spin-flip parts of the SOC. 
This terminology refers to the effect of acting with each of them on an eigenstate of $S_{\hat{n}}$. 
Accordingly, we define $\sigma^{\uu}$ and $\sigma^{\ud}$ as the AHC calculated from Eq.~(1) after 
selectively removing the second or the first term on the right-hand side of Eq.~(2). This is not an exact decomposition, 
but inspection of Table~\ref{table:aahe-3dpt} shows that it is approximately valid for both magnetization
directions in FePt and CoPt, and NiPt
with $\mathbf{M}$ along $x$,~i.e.~$\sigma^{\rm tot} \approx \sigma^{\uu}+\sigma^{\ud}$ in these cases, 
while even for NiPt with $\mathbf{M}\Vert z$ a large discrepancy between the $\sigma^{\rm tot}$
of $-$1165~S/cm and the sum $\sigma^{\uu}+\sigma^{\ud}\approx-2000$~S/cm does not
change the general line of argument we are to present below. 

\begin{table}[ht!]
\begin{ruledtabular}
\begin{tabular}{cc|rr|rr|rr}
\multirow{2}{*}{} &                   & \multicolumn{2}{c|}{SOC $3d$+Pt}  & \multicolumn{2}{c|}{SOC $3d$} & \multicolumn{2}{c}{SOC Pt} \\ 
                  &                   & $\uparrow$  &$\downarrow$    & $\uparrow$ &$\downarrow$      &$\uparrow$   &$\downarrow$   \\ \hline
\multirow{2}{*}{FePt}
       &  [001]  & 612    &  $-$35    &  12   &    $-$1     &  579   &  $-$49                      \\
       &  [110]  & 719    & $-$134    &  35   &   225     &  666   &  $-$282            \\ \hline
\multirow{2}{*}{CoPt}
       &  [001]  & 603    &  $-$98    &  14   &   624     &  588   &  $-$802            \\
       &  [110]  & 728    & $-$241    &  48   &   400     &  661   &  $-$487            \\ \hline
\multirow{2}{*}{NiPt}
       &  [001]  & 1048   & $-$2562   & 87    &  $-$1607    & 1032   & $-$2352           \\
       &  [110]  & 1589   & $-$1879   & 215   &  $-$1461    & 1501   &  $-$1581         \\
\end{tabular}
\end{ruledtabular}
\caption{Spin-resolved contributions to the spin-conserving AHC in $3d$Pt alloys.
SOC $3d$ (Pt, $3d$+Pt) stands for the values obtained with SOC on only $3d$ (Pt, both $3d$ and Pt) 
site(s) included in the calculations. $\uparrow$ ($\downarrow$) denotes the contribution from the 
majority (minority) spin channel. All values are in S/cm. \label{table:decomp2} 
}
\end{table}

By analyzing Table~\ref{table:aahe-3dpt} we observe that the spin-flip conductivity in $3d$Pt alloys provides
a significant contribution to the total AHC, which is particularly striking in case of CoPt where $\sigma^{\ud}$
is even somewhat larger than the spin-conserving part for both magnetization directions. 
And while in NiPt the AHC anisotropy $\Delta\sigma^{\rm tot}$ is mainly given by the anisotropy of its spin-conserving
part, $\Delta\sigma^{\uu}$, in FePt and CoPt the anisotropy of the AHC is driven entirely by the anisotropy of
$\sigma^{\ud}$, $\Delta\sigma^{\ud}$, which exceeds as much as 90\% of $\Delta\sigma^{\rm tot}$ in FePt.
Such a pronounced role of the spin-flip SOC for the anomalous Hall conductivity and its anisotropy in ferromagnets 
containing heavy elements, such as Pt, was demonstrated and explained by Zhang and co-workers by employing
the perturbation theory arguments.\cite{Zhang:2011} In the case of considered here alloys, from Table~\ref{table:aahe-3dpt} it
is however clear that, despite a large spin-flip contribution, the overall trend of the total AHC between FePt and 
NiPt can be qualitatively described by considering the spin-conserving AHC only, and we dedicate the rest of 
the paper to the analysis of $\sigma^{\uu}$ in FePt, CoPt and NiPt compounds. 

Firstly, the advantage of considering exclusively the spin-conserving SOC is that spin remains a good
quantum number, and the conductivity can be unambiguously
decomposed into spin-up and spin-down parts: $\sigma^{\uu}=\sigma^{\uparrow}+\sigma^{\downarrow}$.
In particular, this means that if we assume that in the system considered the spin-orbit is given 
only by the spin-conserving part, the 
corresponding spin Hall conductivity $\sigma_{\rm SH}$ can be obtained as the difference between the 
spin-resolved conductivities: $\sigma_{\rm SH}=\sigma^{\uparrow}-\sigma^{\downarrow}$,\cite{Hirsch:1999}
which implies that in a non-magnetic material, such as~e.g.~Pt, $\sigma^{\uparrow}=-\sigma^{\downarrow}$, 
while $\sigma_{\rm SH}=-2\sigma^{\downarrow}$. Secondly, among the $\sigma^{\uu}$ and $\sigma^{\ud}$
conductivities, the later one is much more sensitive to the details of the Fermi surface, while the electronic 
transitions contributing to the spin-conserving AHC according to Eq.~(1) are distributed much broader in 
energy around $E_F$, \cite{Zhang:2011} which makes the analysis of the latter easier.

In order to get additional insight into the structure of $\sigma^{\uu}$, we use the
atomic decomposition of the AHC for each spin channel, considered by Zhang {\it et al.},\cite{Zhang:2011} based on the following 
atomic decomposition of the spin-orbit part of the Hamiltonian:
\begin{equation}
\label{eq:soc1}
H_{\rm SO}   =\xi_{\text{$3d$}}\mathbf{L}^{3d}\cdot\mathbf{S}+\xi_{\text{Pt}}\mathbf{L}^{\rm Pt}
\cdot\mathbf{S}, 
\end{equation}
where $\mathbf{L}^{3d({\rm Pt})}$ is the orbital angular momentum operator
associated with $3d$ (Pt) atoms, and $\xi_{3d{\rm (Pt)}}$ is the spin-orbit coupling strength 
averaged over the valence 
$d$-orbitals inside $3d$ (Pt) atom, with the values of 0.54~eV for Pt and 0.05$-$0.07~eV for $3d$ transition-metal
atoms. By selectively turning off the spin-orbit coupling inside $3d$ transition-metal atoms ($\xi_{3d}=0$)
or Pt atoms ($\xi_{\rm Pt}=0$) we obtain the values of $\sigma^{\uparrow(\downarrow)}_{\rm Pt}$ and
 $\sigma^{\uparrow(\downarrow)}_{3d}$, respectively.
  
\begin{figure}[t!]
\includegraphics[width=7.6cm]{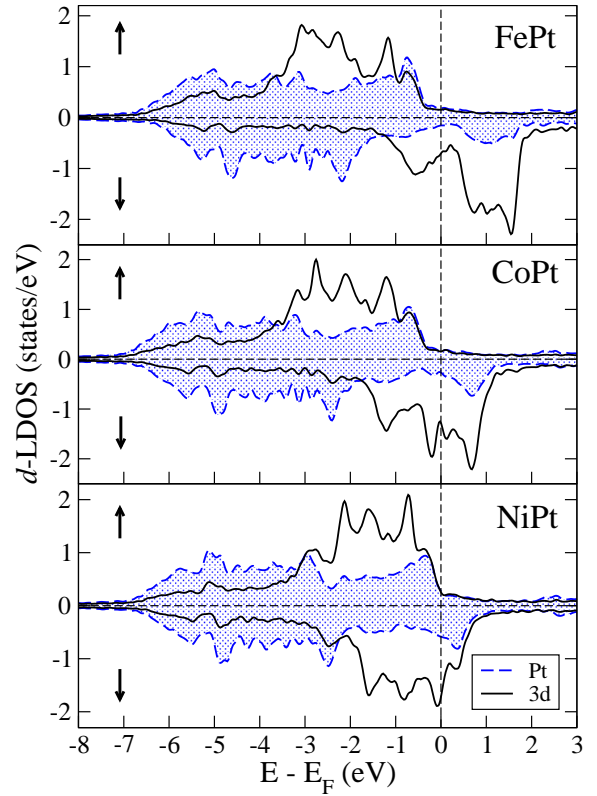}
\caption{
Atomically-resolved density of $d$-states in FePt, CoPt and NiPt alloys. Up- and down-arrows stand for 
spin-up and spin-down. 
\label{fig3}}
\end{figure} 

The results of our calculations for the spin and atomically decomposed $\sigma^{\uu}$ in FePt, CoPt
and NiPt alloys are presented in Table~\ref{table:decomp2}. Let us take a look at the first two columns
of the table, where the values of the total $\sigma^{\uparrow}$ and $\sigma^{\downarrow}$ are listed.
Firstly, we observe that positive $\sigma^{\uparrow}$ and negative $\sigma^{\downarrow}$ are opposite 
in their sign for all alloys. 
Secondly, upon going from FePt to NiPt, the spin-up AHC increases but retains its order of magnitude, 
being about 650~S/cm for FePt and 1300~S/cm for NiPt. On the other hand, a very small spin-down AHC
of $\approx$$-$100~S/cm for FePt increases by an order of magnitude and reaches as much as $-$2500~S/cm
in NiPt. Correspondingly, in FePt the positive sign of $\sigma^{\uu}$ is due to the AHC in spin-up channel,
while in NiPt the negative $\sigma^{\uu}$ is driven by large and negative spin-down AHC. 

Consider now the case of FePt. The atomic decomposition of the AHC, presented in Table~\ref{table:decomp2},
clearly reveals that the large spin-up AHC in this alloy originates from the spin-up contribution of Pt atoms, while 
the Fe contribution to $\sigma^{\uparrow}$ is very small. In the spin-down channel, Pt and Fe AHCs, both with the magnitude of about 200~S/cm, are opposite in sign and suppress each other. In CoPt the spin-up Pt and Fe AHCs remain basically
the same compared to FePt, while the corresponding spin-down conductivities significantly increase in their 
magnitide. This can be related to the increase in both Co and Pt density of states of $d$-electrons around the Fermi 
energy for minority spin which can be clearly seen in Fig.~\ref{fig3}, as compared to respective DOS of FePt alloy.
Such an enhancement of the DOS around $E_F$ results in more occupied and unoccupied $d$-states and corresponding
transitions accross the Fermi energy, which contribute to the AHC according to Eq.~(1) $-$ similar to the situation we came 
accross when analyzing the tight-binding model results previously. While in the latter case the variations of the Fermi
energy in the region of increased DOS resulted in large changes of the AHE, in the case of a complex ferromagnet with
many bands at $E_F$ in which the AHE is not driven by a single band degeneracy, it seems reasonable to assume that
the increased number of available transitions will lead to a larger magnitude of the AHC. The increased
Co and Pt spin-down AHCs are still opposite in sign however, which still suppresses the total $\sigma^{\downarrow}$,
although its value is also somewhat enhanced compared to FePt, which leads to the decrease in overall 
$\sigma=\sigma^{\uparrow}+\sigma^{\downarrow}$ in CoPt,~c.f.~Table~I.  

The reason behind increased $d$-DOS of Pt atoms at the Fermi level in CoPt lies in moving of the spin-down Co
$d$-subband to lower energies with decreasing exchange splitting. This leads to a stronger hybridization between
the Co and Pt $d$-states, which are situated mainly below $E_F$, and increased Pt DOS (Fig.~\ref{fig3}). This
is even more pronounced in case of NiPt, where the spin-down Ni subband lies predominantly below the Fermi
energy and the hybridization with the Pt $d$-states is even stronger (Fig.~\ref{fig3}). Correspondingly, as a 
result of even more enhanced spin-down $d$-DOS of Ni and Pt atoms around $E_F$ in NiPt alloy, the values
of $\sigma^{\downarrow}_{\rm Ni}$  and  $\sigma^{\downarrow}_{\rm Pt}$ become very large, reaching
as much as $-$1600~S/cm for Ni and $-$2300~S/cm for Pt. On the other hand, the increase in Ni and Pt 
AHC in the majority channel is quite moderate due to slightly enhanced DOS, and the total AHC in NiPt becomes 
large and negative. 

The change of sign of the $3d$ spin-down subband AHC between Fe, Co and Ni in $3d$Pt
alloys can be probably related to different orbital character of the $d$-states at the Fermi energy and corresponding
matrix elements of the SOC in these transition metals, which could in turn explain the AHC sign change in elemental
Fe, Co and Ni, observed experimentally, and reproduced from the first principles.\cite{Yao:2004,Wang:2006,Roman:2009,Wang:2007} Such a change of sign as a function of the Fermi level position within the spin subband has been
also demonstrated from first principles calculations of the spin Hall conductivity in Pt,\cite{Guo:2008} as well as from
tight-binding calculations of the spin Hall conductivity in $4d$ and $5d$ transition-metals.\cite{Tanaka:2008} Assuming
that in latter cases the spin-flip contribution to the spin Hall effect is negligible,\cite{Tanaka:2008} this results in a corresponding
sign changes of the $\sigma^{\downarrow}$ Hall conductivity, discussed
previously.

From Table~II we can see that the AHC originating from the Pt atoms is 
generally larger in magnitude than that from $3d$ transition-metal. This can be explained by noticing that the large
spin-orbit constant inside Pt atoms $\xi_{\rm Pt}$  is by an order of magnitude larger than
$\xi_{3d}$. Moreover, a consistently positive and negative sign of large
$\sigma^{\uparrow}_{\rm Pt}$ and $\sigma^{\downarrow}_{\rm Pt}$ throughout the $3d$Pt family
can be related to a small spin-polarization of the Pt atoms,~i.e.~under the condition that this spin-polarization
is at all absent, $\sigma^{\uparrow}_{\rm Pt}$ should be equal to $-\sigma^{\downarrow}_{\rm Pt}$,
while the difference of the both would provide a value of the intrinsic spin Hall conductivity in Pt of about 2000~S/cm.
As we can see from Table~II, these arguments indeed explain the sign and magnitude of the Pt-originated AHC.   
Finally, we would like to remark that although the atomic decompositon for the spin-resolved conductivities 
is overall rather reasonable in that $\sigma^{\uparrow}\approx\sigma^{\uparrow}_{\rm Pt}+\sigma^{\uparrow}_{3d}$
and  $\sigma^{\downarrow}\approx\sigma^{\downarrow}_{\rm Pt}+\sigma^{\downarrow}_{3d}$, see Table~II,
such a decomposition works much better for the majority channel. We attribute this observation to a much stronger
hybridization of the $3d$ and Pt $d$-states for the minority-spin around the Fermi level, which enhances the 
contributions to the AHC for which the presence of SOC on both Pt and $3d$ transition-metal atoms is important.
Such contributions are omitted in the atomic decomposition used above.

\section{Conclusions}

In conclusion, we investigated from the first principles the intrinsic anomalous Hall effect in $3d$Pt alloys.
From our calculations it follows that the AHC in this type of compounds is strongly anisotropic. We demonstrate
the generality of such anisotropy in uniaxial ferromagnets by considering a simple three-band tight-binding
model. In combination with the sign change of the conductivity upon going from FePt to NiPt the pronounced
AHC anisotropy leads to the occurrence of the colossal anisotropic AHE and anti-ordinary AHE in the vicinity of 
the CoPt alloy. While in the case of colossal anisotropic AHE the anomalous Hall current completely vanishes for 
one of the magnetization directions in the crystal, within the scope of the anti-ordinary AHE the rotational sense
of the Hall current is opposite to that of the magnetization, and a complete collinearity of the two can be 
achieved for a certain "magic" angle of the magnetization in the crystal. We relate the general trend of the AHC
in these alloys to the changes in their electronic structure in the vicinity of the Fermi level, and discuss
possible applications of the anisotropic AHE in these compounds.

\section{Acknowledgements}

We acknowledge discussions with Ivo Souza and Frank Freimuth. This work was supported by the HGF-YIG Program
  VH-NG-513. Computational resources
  were provided by J\"ulich Supercomputing Centre.

\end{document}